\documentclass[fleqn,usenatbib]{mnras}

\usepackage[T1]{fontenc}
\usepackage{ae,aecompl}

\usepackage{graphicx}	
\usepackage{amsmath}	
\usepackage{amssymb}	

\title[Optical continuum of stellar superflares]{Soft X-ray heating as a mechanism of optical continuum generation in solar-type star superflares}

\author[B. A. Nizamov]{
Bulat A. Nizamov,$^{1, 2}$\thanks{E-mail: nizamov@physics.msu.ru}
\\
$^{1}$Sternberg Astronomical Institute, M.V.Lomonosov Moscow State University, Universitetskii pr., 13, Moscow 119234, Russia
\\
$^{2}$Faculty of Physics, M.V.Lomonosov Moscow State University, Leninskie Gory, Moscow 119991, Russia
}

\date{Accepted XXX. Received YYY; in original form ZZZ}

\pubyear{2019}

\begin{document}
\label{firstpage}
\pagerange{\pageref{firstpage}--\pageref{lastpage}}
\maketitle

\begin{abstract}
Superflares on the solar type stars observed by \textit{Kepler} demonstrate the contrast
in the optical continuum of the order 0.1--1 per cent. The mechanism of formation of this radiation
is not firmly established. We consider a model where the stellar atmosphere is irradiated
by the soft X rays emitted from the flaring loop filled with the hot plasma. This radiation
heats a large area beneath the loop. Subsequent cooling due to {\sc h$^{-}$} and hydrogen free-bound
emission can contribute to the observed enhanced continuum. We solve the equations of
radiative transfer, statistical equilibrium, ionization balance and radiative equilibrium
in the model atmosphere illuminated by the soft X rays, compute the temperature and
the electron density in the atmosphere and find the emergent radiation. We found that a flare loop
of the length $\sim10^{10}$~cm and plasma density $10^{12}$~cm$^{-3}$ at the temperature
$T = 20$ MK can provide the contrast in the \textit{Kepler} bandpass of 0.1 and 0.8 per cent
if the heated region covers $\sim$~1 and 10 per cent of the visible stellar surface respectively.
The required emission measure is of the order $10^{55}$~cm$^{-3}$.
\end{abstract}




\section{Introduction}
Flaring stars are observed across HR diagram \citep{Pettersen1989, Balona2015}. Among main sequence stars, flares
are most commonly observed on UV Cet type stars which are M dwarfs with emission lines (dMe type).
Flares on hotter stars, in particular G stars, have been reported more rarely. This can be attributed
to the fact that, on brighter stars, only sufficiently powerful flares can be detected. Moreover,
such stars probably flare less frequently.

Observations of flares on G stars were hugely enriched by the \textit{Kepler} mission \citep{Borucki2010}
which continuously observed more than 150000 stars for four years with exceptional photometric
accuracy\footnote{The main mission was succeeded by the \textit{K2} mission \citep{Howell2014} which is still
going on.}. \citet{Maehara2012} reported on 365 superflares on 148 solar-type stars observed in 120
days. \citet{Shibayama2013} extended this result to 1547 flares.
The spectral band of \textit{Kepler} is from 4000 to 9000 \AA{} and the contrast of the
flares in such a wide range (essentially, in white light) is itself striking: the typical value is
1 per cent of the average non-flaring brightness, but sometimes it can reach even 10 per cent\footnote{By contrast, we mean
the quantity $\int R(\lambda)F_f(\lambda)d\lambda/\int R(\lambda)F_q(\lambda)d\lambda - 1$ where $R(\lambda)$
is the \textit{Kepler} bandpass, $F_f$ and $F_q$ are the stellar fluxes during a flare and in the quiescence
respectively.}. Solar white light
flares can provide a contrast of only 0.01 per cent when integrated over the disc \citep{Haisch1991}.
On the other hand, flares on M dwarfs can easily increase the brightness of the star by many times,
e.g., two flares on UV Cet had $U$-band contrast of 8.2 and 127.1 \citep{Bopp1973}, a flare on EV Lac
had $\Delta U=3.62$, i.e. the $U$-band contrast 28 \citep{Alekseev1994}. \citet{Pettersen2016} observed
a $\Delta U=7.2$ flare on EV Lac and \citet{Hawley1991} observed a $\Delta U=4.5$ flare on AD Leo.
See also multiwavelength observations of a giant flare on CN Leo by \citet{Fuhrmeister2008} and a
comprehensive overview by \citet{Gershberg2005}.

However, M dwarfs differ
from solar-type stars in that they are fainter (hence flares are relatively more prominent), smaller
(hence flare photospheric filling factor can be larger) and have somewhat different structure of
the atmosphere due to lower $T_\mathrm{eff}$ and larger $\log g$. Therefore one can ask whether and to what extent the mechanisms
of radiation are common for superfalres on dMe and solar-type stars. As \citet{Heinzel2018} noted in
regard to the results of \citet{Maehara2012}, `only a limited attention was devoted to understanding
the mechanisms of the superflare emission'.

Since the beginning of the study of solar and stellar flares, there have been proposed a number of
mechanisms of heating the lower atmosphere which can result in the optical continuum emission.
\citet{Najita1970} considered the heating of the solar atmosphere by non-thermal electrons and
protons. \citet{Aboudarham1989} argued that the white light emission in solar flares results from heating the
photosphere by the enhanced radiation of the chromosphere which in turn is heated by non-thermal
electron beams. Analogous conclusions are drawn by \citet{Machado1989}, but they argue that the
chromospheric heating by protons is more preferable. Direct heating of the upper photosphere by
non-thermal electrons and protons in case of stellar flares was discussed by
\citet{Grinin1977, Grinin1988, Grinin1989}. Heating of the stellar photosphere by soft X rays from the
flare loop was considered by \citet{Mullan1977} and, more thoroughly, by \citet{Hawley1992} (hereafter HF92).
\citet{Livshits1981} calculated the gasdynamic response of the atmosphere to the bombardment by
non-thermal electrons and proposed that the so-called low-temperature condensation can account
for the optical continuum of stellar flares. The model was refined by the authors in a subsequent paper
\citet{Katsova1997}; the radiation hydrodynamics approach was also developed by
\citet{Fisher1985, Abbett1999, Allred2005}. \citet{Mullan1976} proposed that the photosphere can
be heated by the thermal conduction from the flaring loop filled with the hot plasma. \citet{Heinzel2018}
argue that the continuum radiation of stellar superflares can be dominated by the flare loop
radiation, especially hydrogen free-free radiation.

The mechanisms mentioned above are usually invoked to explain either impulsive or gradual phase
of the flare. The light curves of the flares reported by \citet{Maehara2012} show gradual decay.
In fact, these observations were conducted in the \textit{Kepler}'s long cadence mode (30 min
time resolution), so each flare containing several data points lasted for several hours and was
gradual in nature (although the presence of an impulsive phase cannot be excluded) and the most
energy in the optical range was apparently emitted in the gradual phase. In this paper we will consider
heating of the photosphere by the soft X rays from the flare loop as a possible cause to the optical
continuum to emerge in the gradual phase of superflares on solar-type stars. As was noted above,
this mechanism has been discussed in a number of papers, but, in case of stellar flares, to our
knowledge, they were always assumed to occur on M dwarfs. On the other hand, it was also considered
in the context of solar flares, but for flare energies typical for the Sun. Our aim is to apply
strong heating by the soft X rays to an atmosphere of a solar-type star and find out the
parameters of the hot plasma which can account for the observed contrast of G-type star supeflares
in the \textit{Kepler} bandpass. 
\section{Problem setup}
The general picture of stellar flares is qualitatively analogous to the standard model of
solar flares. First, the energy of the non-potential magnetic field is rapidly released.
Some large portion of this energy is thought to be transferred to non-thermal particles,
electrons or protons. The particles travel along the magnetic field lines and impact the
chromosphere. This process is usually thought to account for the impulsive phase of a flare.
The impact results in a strong heating and evaporation of the chromospheric material
which expands drastically at the temperature of $10^7 - 10^8$~K and fills the magnetic loop.
Thus, the loop becomes a source of soft X rays (SXR) which irradiate the stellar surface.
Our aim is to find the properties of the SXR plasma which could provide irradiation of a
solar-type star so that the disc-integrated brightness in the white light increased by
$\sim 1$ per cent. We do this by solving the set of the equations of radiative transfer, statistical
equilibrium, ionization balance and radiative equilibrium in a stellar atmosphere irradiated
by the SXR. Since we are interested in flares on solar-type stars, we use the Harvard
Smithsonian reference atmosphere \citep{Gingerich1971}, or HSRA, which is a semi-empirical model of
the solar photosphere and the chromosphere up to the height of 1850 km above the point
$\tau_{5000}=1$.

The irradiation by SXR is included as the upper boundary condition in the equations of
radiative transfer. The intensity of SXR is computed with the {\sc chianti} package
\citep{DelZanna2015}.

The equations of statistical equilibrium are solved with the method of accelerated lambda
iteration as implemented by \citet{Rybicki1992}, hereafter RH92. The equation of radiative equilibrium in
its simplest form reads
\begin{equation}
    4\pi\int(\chi_\nu J_\nu - \eta_\nu)d\nu = 0
\end{equation}
where $\chi_\nu$ is the opacity, $\eta_\nu$ is the emissivity and $J_\nu$ is the mean
(i.e. angle averaged) intensity. However, as discussed by \citet{Kubat1999}, this form
can be not appropriate in the outer layers of the atmosphere because of decoupling of
the radiation from the thermal energy reservoir of the atmospheric material. One possible
solution proposed by \citet{Kubat1999} is to use the so called thermal electron balance,
i.e. to write the energy conservation law as applied to the thermal energy of free
electrons rather than to the energy of the radiation. However, when we applied this method
in the case of strong external heating by SXR, we could not reach convergence. Therefore,
we used the equation of radiative equilibrium. To avoid the difficulty of strong line scattering
mentioned above, we applied the technique of the approximate lambda operator in a manner
similar to that of RH92, see appendix \ref{app} for details.

The whole set of equations is solved iteratively in the following way.
\begin{enumerate}
    \item The atomic level
populations and electron number densities at all the depths were assigned LTE values and
the formal solution of the equation of radiative transfer was found.
    \item With the
radiation intensity, electron number density and the temperature fixed, the atomic
level populations are found iteratively following the procedure of RH92.
    \item All the equations (statistical equilibrium, ionization
balance and the radiative equilibrium) are linearized with respect to the
corresponding variables (atomic level populations, electron number density and temperature)
and the corrections are found. However, the corrections to the level populations are
discarded at this step and only electron number density and temperature are corrected.
Such a strategy was reported to have a stabilizing effect by \citet{Mihalas1970, Auer1969},
see also \citet{Kubat1997}.
    \item With the temperature and electron density updated, opacity
is recalculated and a new formal solution for the radiation is found.
    \item Move to the step (ii).
\end{enumerate}
This algorithm was repeated until the corrections of the level populations, electron
number density and temperature were small.

In our calculations we included a 6 level + continuum \ion{H}{i} atom, a 5 level + continuum
\ion{Ca}{ii} atom and a 6 level + continuum \ion{Mg}{ii} atom. The oscillator strengths for hydrogen
were taken from \citet{Wiese2009}, the bound-free opacity was calculated according to
the Kramers formula with the Gaunt factors taken from \citet{Karzas1961}. The free-free
opacity for hydrogen was calculated with the classical formula and a constant Gaunt factor.
For \ion{Mg}{ii} and \ion{Ca}{ii}, oscillator strengths were taken from the VALD database \citep{Ryabchikova2015}
and the photoionization cross sections from the TOPbase project \citep{Cunto1992, Cunto1993}.
Free-free opacity for \ion{Ca}{ii} and \ion{Mg}{ii} was calculated analogously to \ion{H}{i}. This rough approximation
is applicable because free-free processes for these atoms contribute very little to the overall opacity.
The abundance of the negative hydrogen ion was calculated in LTE relative to the ground
state of neutral hydrogen. Bound-free and free-free opacities for it were calculated
according to \citet{Kurucz1970}.

In this work we only took into account collisions with electrons. Electron impact excitation
rates were taken from \citet{Przybilla2004} for \ion{H}{i}, \citet{Melendez2007} for \ion{Ca}{ii} and
\citet{Sigut1995} for \ion{Mg}{ii}. The data for the impact ionization were taken from
\citet{Mihalas1967} for \ion{H}{i} and the Seaton formula was used for \ion{Ca}{ii} and \ion{Mg}{ii} \citep{Seaton1962}.

When calculating the electron density from the charge conservation equation, we take into
account that the ionization of \ion{H}{i}, \ion{Mg}{ii} and \ion{Ca}{ii} obeys the equation of statistical equilibrium;
also, we include the LTE ionization of the neutral helium, carbon, oxygen, silicon and iron.

The SXR heating is enabled via ionization of hydrogen and K-shell photoionization of
helium, carbon, oxygen, magnesium and silicon. The absorption cross sections in this spectral
range are taken from \citet{Karzas1961} for hydrogen and computed with the code
{\sc mucal}\footnote{\url{http://ixs.csrri.iit.edu/database/programs/mcmaster.html}}
which uses the compilation of X-ray cross sections from \citet{McMaster1969} for heavy atoms.
While the ionization of hydrogen is treated self-consistently, this is not the case for
heavier atoms: we did not take into account the cascade transitions after K-shell
ionization and corresponding radiative losses. Therefore, when computing the heating
due to K-shell ionization of non-hydrogenic atoms, we only took into account the energy
of the escaping electron, so that the term in the energy balance equation corresponding
to the K-shell ionization reads
\begin{equation}
    Q_\mathrm{SXR} = 4\pi n \int_{h\nu_0}^\infty \alpha_\nu J_\nu (1 - \frac{h\nu_0}{h\nu}) d\nu
\end{equation}
where $n$ is the number density of atoms of a certain element, $h\nu_0$ is the threshold
of the K-shell ionization, $\alpha_\nu$ is the ionization cross section.

The intensity of the SXR was calculated in the {\sc chianti} package. We chose the constant
temperature of plasma to be $T=20$ MK which is quite typical for stellar flares.

Three values of the emission measure were tried: $\log\mathrm{EM} = 33, 34, 34.5$.
Note that {\sc chianti} computes the \textit{column} emission measure, 
which implies that in our model calculations, the source of SXR is represented by a box on
the top of the model atmosphere homogeneously filled with the hot plasma. The height of
the box $h$ and the plasma density in it $n$
are such that $n^2 h = \mathrm{EM}$ where EM has the values indicated above. The horizontal
extent of the box should be chosen so that the corresponding volume emission measure is
comparable to the observed values. Our important assumption is that, if the plasma within the box is
condensed into a flare loop (conserving the temperature and the volume emission measure),
the irradiation and heating of the stellar surface below it does not change radically.
The relation between the box size and the flare loop parameters is discussed in the next section.

HF92 found that the X rays with the wavelengths \mbox{$\gtrsim 25$~\AA}{} are absorbed
in the highest levels of the chromosphere. In our model, the chromosphere extends up to
1850~km and we also see that the longer wavelength X rays are absorbed at the highest
levels. These levels do not contribute to the heating of the photosphere and do not
produce considerable optical continuum. Moreover, strong heating of these layers makes
the calculation unstable so that it can fail to converge. Therefore we assumed the SXR
emission to span from 0.1 to 25~\AA. Since the high resolution spectrum in this region
is not of much importance to us, we smoothed it with an instrumental profile of FWHM = 3~\AA{} to keep the frequency
grid relatively coarse.

In principle, our method of solution is close to that of HF92 (except that we do not model
the flaring loop itself and the transition region). The important difference is that we apply the strong
heating to the solar atmosphere model and assume that it is not only the footpoint of
the flaring loop which is heated by SXR, but a large area below the loop. Also, we use more detailed SXR spectrum
and solve the equation of radiative equilibrium within the global iterative procedure.
\section{Results and discussion}
Using the techniques described in the previous section, we calculated the temperature and
electron density in the atmosphere heated by the external SXR. We should note that we did
not solve the equation of hydrostatic equilibrium and the heavy particle stratification
was fixed. We assume that this simplification does not affect our results significantly.
The temperature and electron density in the heated models are shown in Figs.~\ref{fig:temperature}
and \ref{fig:ne}. In these and following figures, $m$ is the mass column density in g cm$^{-2}$.
\begin{figure}
	\includegraphics[width=\columnwidth]{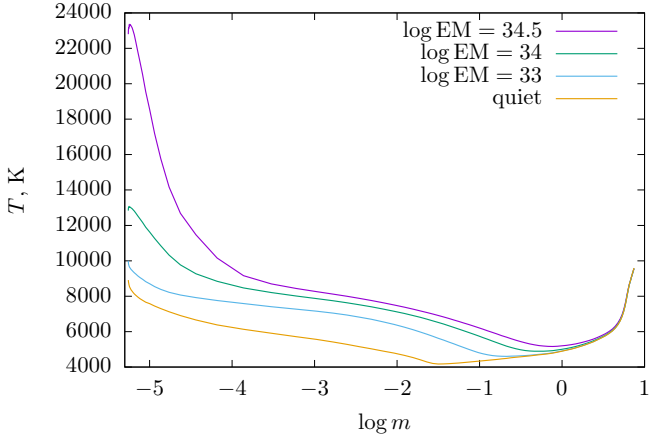}
    \caption{The temperature in the quiescent model and heated models}
    \label{fig:temperature}
\end{figure}
\begin{figure}
	\includegraphics[width=\columnwidth]{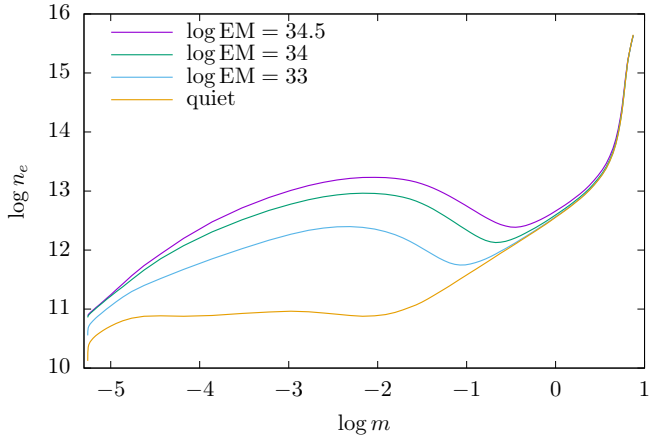}
    \caption{The electron density in the quiescent model and heated models.}
    \label{fig:ne}
\end{figure}
One can see that all the three models can provide quite strong heating below
the temperature minimum, but the maximum column depth where heating is present increases with
the emission measure of the hot plasma. In the following, we will refer to the results obtained
for the strongest heating ($\log \mathrm{EM}=34.5$) as the most representative case.

\begin{figure}
	\includegraphics[width=\columnwidth]{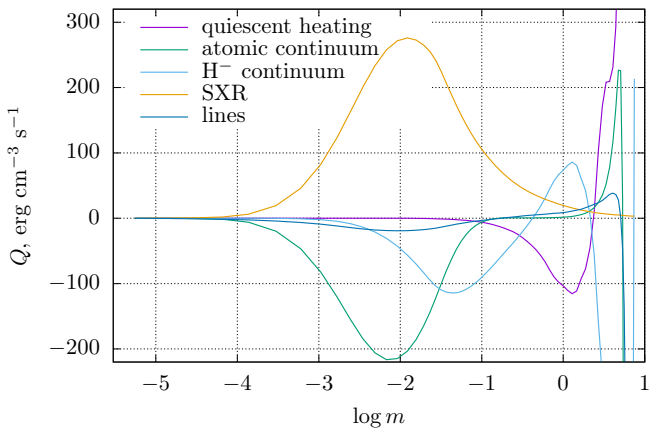}
    \caption{The contributions of various atomic processes to the overall heating (positive values)
    and cooling (negative values) in the model atmosphere.}
    \label{fig:heating}
\end{figure}

Fig.~\ref{fig:heating} shows the processes responsible for the heating and cooling of the
irradiated atmosphere.
Positive values correspond to heating and negative ones to cooling.
The heating by SXR has a broad maximum around $\log m = -2$. Note however
that the $x$ scale is logarithmic; in linear scale, the SXR heating shows abrupt rise and much slower
decline with the depth. The heating shown in Fig.~\ref{fig:heating} is calculated per unit volume.
Therefore, although the heating at the top of the model atmosphere is low, the temperature rise
in this region is high due to the low density as can be seen in Fig.~\ref{fig:temperature}.
Recall once again that we restricted the SXR spectrum by the limits 0.1 -- 25~\AA. Radiation
at longer wavelengths would be absorbed in the highest levels of the atmosphere which would result
in even stronger heating. Our test calculations in case of low emission measure show that
this increased heating of the outer layers has small effect on the temperature of the
underlying material and the overall amount of the optical continuum.

The SXR heating is balanced by the cooling due to the line emission, atomic continuum and {\sc h$^{-}$}
continuum. In case of relatively weak heating ($\log \mathrm{EM} = 33$), the cooling is
realized by the line emission (peaks at $\log m = -2.2$) and {\sc h$^{-}$} emission (peaks at
$\log m = -1.2$), while the contribution from the atomic continua is weak. When the emission
measure and SXR heating increase, the cooling rate due to {\sc h$^{-}$} increases accordingly,
while the line cooling grows slowly and cooling due to atomic continuum processes becomes more
important. In case of $\log \mathrm{EM}=34.5$ (Fig.~\ref{fig:heating}) line cooling is about one
tenth of atomic continuum cooling. The main components of this continuum are Balmer and, to a
lesser extent, Paschen continua.
The cooling feature arises because, due to the enhanced temperature, spontaneous bound-free
emission dominates over the photoabsorption. In deeper layers, the temperature goes down,
and both hydrogen second level population and ionization degree become lower, so that both
heating and cooling decrease. In these layers cooling is due to {\sc h$^{-}$} bound-free emission.

HF92 found that the atmosphere is heated not only directly by the SXR, but also by the
hydrogen free-bound emission arising from the regions heated by SXR, the effect predicted
for the Sun by \cite{Avrett1986}. In our calculations, we do not see significant heating
below the temperature minimum region due to hydrogen free-bound emission. This heating is
still caused by the shortest wavelength SXR penetrating to such deep layers.

In figure~1 of HF92, one can see that the quiescent heating is not zero, but has a minimum
around $\log m = 0$ followed by a larger maximum. The authors attribute this maximum to the
fact that the model they used included convective heating, while they took into account
only radiative heating and cooling. Our model has similar features, but they are larger in
absolute value. We suppose that the interpretation of HF92 is applicable in our case since
\cite{Gingerich1971} tried to take the convection into account when constructing HSRA.

\begin{figure}
	\includegraphics[width=\columnwidth]{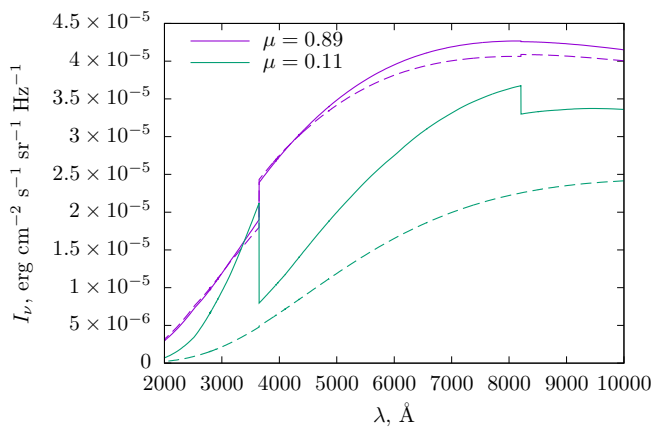}
    \caption{The continuum spectra of the radiation from the heated model atmosphere for
    the two viewing angles. Dashed lines represent the quiescent spectrum. The hot plasma
    emission measure is $\log\mathrm{EM} = 34.5$.}
    \label{fig:spectra}
\end{figure}

\begin{figure}
	\includegraphics[width=\columnwidth]{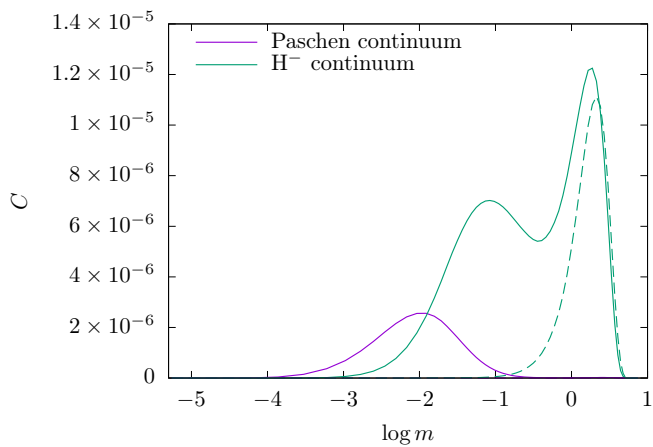}
    \caption{Contribution functions for Paschen and {\sc h$^{-}$} free-bound emission. Dashed lines
    correspond to the quiet atmosphere. The hot plasma emission measure is
    $\log\mathrm{EM} = 34.5$ and the viewing angle cosine is $\mu = 0.11$}
    \label{fig:contribution}
\end{figure}

External heating of the atmosphere affects the emergent radiation spectrum. We found that
the contrast, i.e. the ratio of the emergent intensity from the heated atmosphere to that
of the quiescent atmosphere, strongly depends on the viewing angle. In other words, it
depends on the location of the heated region on the stellar disc: in the center of the disc,
the contrast is smaller while near the limb it is larger. Spectra of the heated and quiescent
models for the two locations of the flaring region are shown in Fig.~\ref{fig:spectra}.
We computed the contribution functions for Paschen continuum and {\sc h$^{-}$} free-bound emission
at the reference wavelength of 5000~\AA{} according to \cite{Magain1986}:
\begin{equation}
C = \frac{\ln 10}{\mu}\frac{\eta}{\chi}\tau_{5000}e^{-\tau_{5000}/\mu}
\end{equation}
where $\eta$ is the {\sc h$^{-}$} or Paschen free-bound emissivity, $\chi$ is the total absorption
(both taken at 5000~\AA), $\tau_{5000}$ is the optical depth at 5000~\AA, $\mu=\cos\theta$ is
the cosine of the viewing angle. Fig.~\ref{fig:contribution} shows the contribution functions
for the more representative case of $\mu=0.11$ (close to the limb).
Dashed lines correspond to the quiescent atmosphere.
One can see that Paschen continuum does contribute to the overall intensity, but the main
contribution is due to {\sc h$^{-}$} and comes from deeper layers. However, the significant enhancement
of the {\sc h$^{-}$} emission around $\log m = -1$ is located substantially higher in the atmosphere
than the region from which the bulk optical continuum emerges. This allows to explain the
dependence of the contrast on the viewing angle on the basis of the Eddington-Barbier relation [see
\citet{Mihalas1978}]. At the disk center we see the bright radiation of the photosphere and
the excess radiation of the heated region is lost in it. When approaching the limb, the photospheric
radiation becomes suppressed and the heated region contributes relatively more to the overall
intensity.

For decades of observations of stellar flares and solar white light flares (WLFs) there has been a
debate about the depth and mechanism of the optical continuum formation. \cite{Avrett1986}
argue that the optical continuum can be due to the hydrogen free-bound emission from the overheated
upper chromosphere. However they note that this same emission can also heat temperature minimum
region and cause the {\sc h$^{-}$} radiation to emerge. The idea of backwarming was used by
\cite{Isobe2007} to interpret the morphology of white light emitting patches in an X class solar flare.
They found that the region of backwarming is located 100 km below (i.e. very close to) the region of the primary heating.
The latter is associated with the region where non-thermal electrons deposit their energy.
In the model atmosphere we used, 20 keV electrons stop at $\log m = -4$ \citep{Syrovatskii1972},
high above the temperature minimum region. 100 keV electrons stop at $\log m = -2.5$, close to the
depth where we obtained the maximum hydrogen continuum cooling. However, the electron spectrum
should be rather hard in order for the bulk energy to be contained in such energetic particles.
Or, alternatively, heating by protons should be invoked.
Note also that the region heated by non-thermal electrons emits in extreme ultraviolet while we considered
irradiation by soft X rays. Therefore the two models are probably not directly comparable.
Also, to interpret the gradual decay phase with the backwarming mechanism, one has to assume
continuous precipitation of the non-thermal particles as is the case in solar two-ribbon flares.
In our model, we assume that there is no sustained heating when the loop is filled with the hot plasma.

In a recent work, \cite{Heinzel2017} report
the observations of a white light emission in an off-limb solar flare coming from the height of
$\sim 1000$~km which they attribute to the Paschen continuum. On the other hand, \cite{MartinezOliveros2012}
report the observations of white light emission from as low as 200--400~km above the photosphere.
Moreover, there are observations of WLFs with weak Balmer continuum \citep{Hiei1982, Boyer1985};
\citet{Fang1995} discuss the two types of solar WLFs supporting either hydrogenic or {\sc h$^{-}$}
origin of the optical continuum. In case of stellar flares, the identification of the optical
continuum origin is even harder. To our knowledge, photometric and spectral observations are
available only for flares on dMe stars [see e.g. \cite{Kowalski2013} and references therein]
and for solar-type stars we only have to rely on the models. In our calculations we see that
the cooling of the region above the temperature minimum (at $\log m \sim -2$) due to the
hydrogen recombination is partly responsible for the optical continuum generation, but the main
contribution is from {\sc h$^{-}$} free-bound radiation emerging from below the temperature minimum
region ($\log m \sim -1$). One should also remember that \textit{Kepler} is completely insensitive
to the Balmer continuum which therefore is excluded from possible causes of the flare contrasts
reported by \cite{Maehara2012}.

As was mentioned above, the contrast in the specific intensity grows from the disc center
to the limb. However, in order for the observed contrast (which is the contrast in the flux)
to be sufficiently large, one has to assume fairly large flare areas. If the heated region has
an area $A$ and is located at an angle $\theta$ from the disc center, its projected area
equals $A\mu$ where $\mu=\cos\theta$. Therefore, when approaching the limb, the flux contrast
drops due to the small projected area of the heated region. We found that the flux contrast is
largest when the heated region is located close to the disc center.

Finally, let us discuss the flaring loop properties. In a recent work, \cite{Guarcello2019}
report simultaneous observations of stellar superflares with \textit{Kepler} and
\textit{XMM Newton}. Apart from a number of M stars, they
observed one F9 star HII~405 and one G8 star HII~345 and the maximum flare emission measure on them were
$\log \mathrm{EM} = 53$ and $\log \mathrm{EM} = 53.6$ respectively. They also estimated the
flare loop length and obtained values of the order $10^{10}$~cm.

Consider a cylindrical volume
of hot plasma with the column emission measure of $3 \times 10^{34}$~cm$^{-5}$. To have the
volume emission measure of $10^{54}$~cm$^{-3}$ its radius should be $3 \times 10^9$~cm which is
close to the radius of a loop inferred from the observations of \citet{Guarcello2019}. With a
reference stellar radius of $7 \times 10^{10}$~cm (the Solar value), such a circular region covers
approximately 0.2 per cent of the visible stellar disk. If we assume such an area of the heated region
in our calculations, we obtain the contrast of only 0.02 per cent whereas \citet{Guarcello2019} observed
the contrast of 1 per cent. However, HII~345 is not a Solar twin: it has a radius of $0.77R_{\sun}$
and the effective temperature $T_\mathrm{eff}=5150$~K, therefore our calculations made for a solar twin
should underestimate the visible contrast. If we assume that the heated region covers 1 per cent or 10 per cent
of the visible disk, the contrast becomes 0.1 per cent and 0.8 per cent respectively (recall that \citet{Maehara2012}
report typical contrast of 0.1--1 per cent). This might imply still larger volume emission measures
than that reported by \citet{Guarcello2019}, up to several times $10^{55}$~cm$^{-3}$.
\citet{Pye2015} observed $\sim 130$ flares on F-M
stars and one can see from their figure~18 that emission measure in excess of $10^{54}$~cm$^{-3}$
is rarely observed. On the other hand, \citet{Aulanier2013} argue that a significant portion of
the stellar surface should be involved in flares of energies $\gtrsim 10^{35}$~erg (see their figure 4).
Recent results from Doppler imaging by \citet{Kriscovics2019} (see also references therein) show that
spots covering order of 10 per cent of the visible surface of young solar analogs are observed.
Moreover, there are observations \citep{Osten2010, Drake2008} which indicate the
presence of the fluorescent iron K~$\alpha$ line which forms when SXR remove an electron from
the inner shell of the iron atom. Figure~8 of \cite{Osten2010} shows that the area illuminated
by SXR is much larger than the footpoints of the flaring loops which are usually supposed to be
the sources of the optical continuum. It is this area which we suppose to be heated by SXR and
produce the optical continuum in solar-type star superflares.

What are the loop parameters which can provide the volume emission measure of $10^{54}$~cm$^{-3}$?
First, the loop size is constrained at least by the size of the star.
\citet{Shibata2002} report loop lengths between $10^{10}$ and $10^{12}$~cm based on
the observations of stellar flares with ASCA and Chandra. \citet{Tsuboi2016} report
loop lengths significantly larger than the solar radius based on the observations with MAXI,
although their results may be overestimation. According to \citet{Namekata2017}, the upper
bound of the loop length is of order $10^{11}$~cm.

Second, the electron density is constrained by the plasma cooling time which equals \citep{VandenOord1989}
\begin{equation}
    \tau = \frac{3 n_e k T}{n_e^2 \Psi(T)}
\end{equation}
with $\Psi(T) = 10^{-24.73}T^{1/4}$~erg~cm$^3$~s$^{-1}$. We see that very high density leads
to very short cooling time which may not be observed. The duration of flares in X rays
and optics reported by \citet{Guarcello2019}
is of order $10^3$ s. \citet{Tsuboi2016} observed a number of flares on stars of various
types with the MAXI instrument and report X-ray duration of several thousands of seconds on
dMe stars. \citet{Pye2015} report the X-ray rise and fall times between $\sim 200$ and $\sim 20000$~s.
Therefore, we can expect the cooling time to be of the order $10^3$~s.

Third, we take into account that the loop can expand upwards. \citet{Schrijver1989} modeled
coronal loops on Capella and $\sigma^2$~CrB. They argue that there exist two populations of
loops (cool and hot) both of which expand towards the apex. The ratio of the cross sections
at the top and the footpoint $\Gamma$ appears to be $\sim 20$ for cool loops and $\sim 4$
for hot loops or $\sim 10$ if the same value holds for both populations. Some researchers
analyzed distributions of the differential emission measure [DEM($T$)] in stellar coronae
and obtained steep dependencies which can be explained with expanding loops \citep{Scelsi2005,Testa2005}.

To summarize the above considerations, let us take the loop length $L = 3 \times 10^{10}$~cm,
the loop cross-section at the footpoint $A_\mathrm{f} = 10^{19}$~cm$^2$, the expansion factor $\Gamma = 5$ and
the electron number density $n_e = 10^{12}$~cm$^{-3}$. Then the volume emission measure
equals \citep{VandenOord1989}
\begin{equation}
    \mathrm{EM} = n_e^2 (\Gamma + 1) A_\mathrm{f} \frac{L}{2} \approx 10^{54}\text{cm}^{-3}.
\end{equation}
and the cooling time $\tau \approx 700$~s which is in accordance with the results of
\citet{Namekata2017}. Moreover, according to \citet{Namekata2017} and \citet{Schrijver1989},
the values of $L$ and $\Gamma$ which we adopted are not extreme and can be further
increased probably providing the values of EM as high as $10^{55}$~cm$^{-3}$.

In order for the hot plasma to be confined inside the loop, the magnetic field should be
sufficiently strong. Equating the gas pressure to the magnetic field energy density, we find
\begin{equation}
    B \gtrsim \sqrt{8\pi \times 2 n_e k T} = 370\text{G}
\end{equation}
which agrees with the data of \citet{Namekata2017} and \citet{Mullan2006}.

A remark on the duration of the white light emission can be made. \citet{Namekata2017} analyzed
a set of solar WLFs and \textit{Kepler} superflares from the viewpoint of the scaling
between the flare energy and the duration of the white light emission. They found that
both sets obey the scaling with the same slope [$\tau \propto E^{1/3}$, this scaling was
earlier used by \citet{Maehara2015}], but with different
prefactors. The authors explain this by finding a new scaling $\tau \propto E^{1/3}B^{-5/3}$
and proposing that the magnetic field in superflare stars is stronger than that in the Sun.
Our analysis cannot confirm or rule out this argument, but we would like to note that the
reconnection timescale used by \citet{Maehara2015} and \citet{Namekata2017} is essentially
the timescale of energy release, i.e. the \textit{heating} timescale. On the other hand, when
analyzing stellar flare observations, different models are used: some models interpret
the decay phase as a cooling process [which can be slowed down by some sustained heating,
see \citep{Reale2007}] which does not depend on the reconnection
while the others assume that the heating is strong enough to
actually control the decay like in the model of two-ribbon flares \citep{Kopp1984, Livshits2002}.
In our model the decay time is attributed to
the radiative cooling. \citet{Landini1986} observed a strong flare in SXR on a young
solar-type star $\pi^1$~UMa and found parameters quite close to ours (they derived smaller emission
measure of $7 \times 10^{52}$~cm$^{-3}$, but the flare energy of $2 \times 10^{33}$~erg is
also rather small for superflares) and they interpreted the flare decay time ($10^3$~s) as due to pure cooling.
Therefore the relation between solar and stellar flare characteristics may not be straightforward.
Probably larger statistics on X-ray observations of superflares on solar-type stars is
needed to understand the result of \citet{Maehara2015} and \citet{Namekata2017}.


Interestingly, the plasma density value of $10^{12}$ -- $10^{13}$~cm$^{-3}$ appears in the work of
\cite{Heinzel2018} who take it as a condition for the optical continuum to emerge directly
from the flare loops due to the hydrogen free-free and free-bound emission. This is supported
by the observations of the famous X8.2 solar flare on 2017 September 10
\citep{Jejcic2018} where the authors interpret the white light emission from the coronal loops by the Balmer and
Paschen continua and hydrogen free-free emission at the electron density of $10^{13}$~cm$^{-3}$.

Superflares are rare on solar-type stars. For example, \citet{Maehara2012} found them on 148 out
of 83000 observed G-type stars. In general,
one should be cautious about the term 'solar-type star'. \citet{Notsu2019} carried
out a spectroscopic study of the superflare stars studied earlier by the group using newer stellar
parameters, in particular stellar radii from \textit{GAIA}-DR2 \citep{Berger2018}. They found
that 40 per cent of the stars reported by \citet{Shibayama2013} are subgiants. The rest 60 per cent, although
being main sequence stars having fundamental parameters close to the solar ones, are in fact
much younger than the Sun and have ages less than 1~Gyr. Most of them rotate much faster than the
Sun [see figure~12 of \citet{Notsu2019}]. \citet{Notsu2019} also found that the maximum spot coverage
of the superflare stars is constant at rotation periods less than 12 days and decreases for slower
rotators. The authors argue that this can be compared with the saturation of the coronal X ray
activity discussed by \citet{Wright2011}. We believe that it is the high level of activity of
young stars that ultimately makes possible flaring plasma parameters which we used in our study.
\section{Conclusions}
We investigated the possibility of the optical continuum generation in \textit{Kepler}
superflares. Our hypothesis is that the hot plasma in the flaring loop irradiates a large
portion of the stellar surface underneath the loop by soft X rays. This leads to the heating
of the lower layers of the atmosphere. The heating is balanced by the cooling due to {\sc h$^{-}$}
and partly hydrogen free-bound emission. This emission creates the continuum contrast which
approaches that observed in the \textit{Kepler} superflares. The hot plasma parameters which
we found to provide significant continuum enhancement are the temperature $T = 20$~MK,
the column emission measure $\log \mathrm{EM} = 34.5$ which we relate to the volume emission
measure of $\log \mathrm{EM} \sim 54-55$ at the density of $10^{12}$~cm$^{-3}$. We assumed that
the region of the lower atmosphere heated by SXR spans up to $\sim 0.1$ of the visible stellar
surface providing the contrast of up to 0.8 per cent.
Such values of the physical parameters have observational support [e.g., \citep{Guarcello2019, Kriscovics2019}]
and discussed by other authors in the context of other possible mechanisms of the optical
continuum formation [e.g., \citep{Heinzel2018}].
\section*{Acknowledgements}
We thank M.M.Katsova and V.P.Grinin for fruitful discussion and the anonymous referee
for valuable comments which helped to improve the paper.

The work was supported by the `BASIS' foundation for the Advancement of Theoretical Physics
and Mathematics and the RFBR grant 19-02-00191. The author acknowledges the support from the
Program of development of M.V. Lomonosov Moscow State University (Leading Scientific School
'Physics of stars, relativistic objects and galaxies').

This work has made use of the VALD database, operated at Uppsala University, the Institute of
Astronomy RAS in Moscow, and the University of Vienna.

CHIANTI is a collaborative project involving the following Universities: Cambridge (UK),
George Mason and Michigan (USA).




\bibliographystyle{mnras}
\bibliography{Nizamov_Optical_continuum} 




\appendix
\section{Radiative equilibrium with approximate lambda operator} \label{app}
Here we show how the formalism developed by RH92 for the equations of statistical equilibrium
can be applied to the equation of radiative equilibrium so that the equation becomes linear
in atomic level populations. Using the expressions for the opacity
$\chi_\nu$ and emissivity $\eta_\nu$ from RH92, one can write the equation of radiative
equilibrium as
\begin{multline}
    4\pi \int(\chi_{\nu}J_{\nu} - \eta_{\nu})d\nu = \\
    = 4\pi \int\left[\sum_{l \succ l'} (n_{l'}V_{l'l} - n_l V_{ll'})J_{\nu} - \sum_{l \succ l'}n_l U_{ll'}\right]d\nu + F = 0 \label{RE}
\end{multline}
where $F$ stands for the free-free processes, {\sc h$^{-}$} emission and absorption and heating by
SXR. The mean intensity is expressed as
\begin{equation}
    J_{\nu} = J_{\nu}^\dagger - \Psi_\nu^* \eta_{\nu}^\dagger + \Psi_\nu^* \eta_{\nu}.
\end{equation}
where the $^\dagger$ symbol means that the quantity is taken from the previous iteration
(the `old' value) and $\Psi_\nu^*$ is the analog of the approximate lambda operator. With this substitution,
equation~(\ref{RE}) reads
\begin{multline}
    4\pi \int\left[\sum_{l \succ l'} (n_{l'}V_{l'l} - n_l V_{ll'}) \left(J_{\nu}^\dagger - \Psi_{\nu}^*\sum_{m \succ m'}n_m^\dagger U_{mm'} + \right.\right. \\
    \left.\left. + \Psi_{\nu}^*\sum_{m \succ m'}n_m U_{mm'}\right) - \sum_{l \succ l'}n_l U_{ll'}\right]d\nu + F = 0.
\end{multline}
As with the equations of statistical equilibrium, we get a nonlinearity due to the `critical'
summation $\sum_{m \succ m'}n_m U_{mm'}$. Analogously to the logic of RH92, we substitute
`new' populations with the `old' ones in the following way:
\begin{multline}
4\pi \int\left[\sum_{l \succ l'} (n_{l'}V_{l'l} - n_l V_{ll'})J_{\nu}^\dagger - \right. \\
\left. - \sum_{l \succ l'}(n_{l'}V_{l'l} - n_l V_{ll'})\Psi_{\nu}^*\sum_{m \succ m'}n_m^\dagger U_{mm'} +
\right. \\ \left.
+ \sum_{l \succ l'}(n_{l'}^\dagger V_{l'l} - n_l^\dagger V_{ll'})\Psi_{\nu}^*\sum_{m \succ m'}n_m U_{mm'} - \sum_{l \succ l'}n_l U_{ll'}\right]d\nu + F = 0.
\end{multline}
Now, using preconditioning within the same transition and taking into account that summations
are over ordered pairs ($m \succ m'$) we can set $m=l, m'=l'$, i.e.
\begin{multline}
4\pi \int\left[\sum_{l \succ l'} (n_{l'}V_{l'l} - n_l V_{ll'})J_{\nu}^\dagger - \sum_{l \succ l'}(n_{l'}V_{l'l} - \underline{n_l V_{ll'}})
\Psi_{\nu}^*n_l^\dagger U_{ll'} +
\right. \\ \left.
+ \sum_{l \succ l'}(n_{l'}^\dagger V_{l'l} - \underline{n_l^\dagger V_{ll'}})\Psi_{\nu}^*n_l U_{ll'} - \sum_{l \succ l'}n_l U_{ll'}\right]d\nu + F = 0.
\end{multline}
One can see that the underlined terms cancel out leaving us with the resulting equation:
\begin{multline}
4\pi \int\left[\sum_{l \succ l'} (n_{l'}V_{l'l} - n_l V_{ll'})J_{\nu}^\dagger - \sum_{l \succ l'}n_{l'}V_{l'l}\Psi_{\nu}^*n_l^\dagger U_{ll'} + \right. \\
+ \left. \sum_{l \succ l'}n_{l'}^\dagger V_{l'l}\Psi_{\nu}^*n_l U_{ll'} - \sum_{l \succ l'}n_l U_{ll'}\right]d\nu + F = 0.
\end{multline}


\bsp	
\label{lastpage}
\end{document}